\documentclass{article}
\usepackage{ijcai24}

\usepackage{times}
\usepackage{soul}
\usepackage{url}
\usepackage[hidelinks]{hyperref}
\usepackage[utf8]{inputenc}
\usepackage[small]{caption}
\usepackage{graphicx}
\usepackage{amsmath}
\usepackage{amsthm}
\usepackage{booktabs}
\usepackage{algorithm}
\usepackage{algorithmic}
\usepackage[switch]{lineno}
\usepackage{xcolor}
\usepackage{array}
\usepackage{amssymb}

\usepackage[inline]{enumitem}
\usepackage{makecell}
\title{When Large Language Models Meet Vector Databases: A Survey}

\newcommand*\samethanks[1][\value{footnote}]{\footnotemark[#1]}

\author{
Zhi Jing*$^1$, 
Yongye Su*$^2$, 
Yikun Han*$^3$, 
Bo Yuan$^4$, 
Haiyun Xu$^6$,\\
Chunjiang Liu$^5$\thanks{Corresponding author.}
Kehai Chen$^4$\samethanks,
Min Zhang$^4$\\
\affiliations
$^1$Carnegie Mellon University\\
$^2$Purdue University\\
$^3$University of Michigan\\
$^4$Harbin Institute of Technology (Shenzhen)\\
$^5$National Science Library (Chengdu), Chinese Academy of Sciences\\
$^6$Shandong University of Technology\\
\emails
zjing2@cs.cmu.edu,
su311@purdue.edu, 
yikunhan@umich.edu, 
23s051006@stu.hit.edu.cn, 
liucj@clas.ac.cn,	
chenkehai@hit.edu.cn,	
xuhy@sdut.edu.cn,
zhangmin2021@hit.edu.cn
}

\begin{document}

\maketitle

\begin{abstract}
This survey explores the synergistic potential of Large Language Models (LLMs) and Vector Databases (VecDBs), a burgeoning but rapidly evolving research area. With the proliferation of LLMs comes a host of challenges, including hallucinations, outdated knowledge, prohibitive commercial application costs, and memory issues. VecDBs emerge as a compelling solution to these issues by offering an efficient means to store, retrieve, and manage the high-dimensional vector representations intrinsic to LLM operations. Through this nuanced review, we delineate the foundational principles of LLMs and VecDBs and critically analyze their integration's impact on enhancing LLM functionalities. This discourse extends into a discussion on the speculative future developments in this domain, aiming to catalyze further research into optimizing the confluence of LLMs and VecDBs for advanced data handling and knowledge extraction capabilities.
\end{abstract}

\section{Introduction}
The field of Natural Language Processing (NLP) achieves a great milestone when ChatGPT unlocks the power of interactive smart assistants and people are offered the unprecedentedly tangible experience of talking to a machine program that is seemingly more knowledgeable than an ordinary person. This remarkable breakthrough is largely attributed to the advancements in LLMs, such as GPT ~\cite{brown2020language,achiam2023gpt}, T5 ~\cite{raffel2020exploring}, and Llama ~\cite{touvron2023llama}. They can process, understand, and generate human-like text. Thanks to the power of pre-training, which involves training a language model on a massive corpus of text data, LLMs can capture the complexities of human languages, including context, idioms, and even cultural references.

However, despite the impressive capabilities of LLMs, they are still under the shadows of doubt in certain aspects.
\begin{enumerate*}
    \item One major shortcoming is the problem of \textbf{hallucination}, where LLMs generate plausible but factually incorrect or faithfully nonsensical information ~\cite{huang2023survey}. There are many potential causes behind this. First, LLMs lack domain knowledge. The fact that LLMs are primarily trained on public datasets ~\cite{penedo2023refinedweb} inevitably leads to a limited ability to answer domain-specific questions that are out of the scope of their internal knowledge. Moreover, real-time knowledge updates are challenging for LLMs. Even if the questions are within the learning corpus of LLMs, their answers may still exhibit limitations because the internal knowledge may be outdated when the outside world is dynamic and keeps changing ~\cite{onoe-etal-2022-entity}. Last but not least, LLMs are found to be biased. The datasets used to train LLMs are large, which may introduce systematic errors ~\cite{10.1145/3442188.3445922}. Essentially, every dataset can be questioned with bias issues, including imitative falsehoods ~\cite{lin-etal-2022-truthfulqa}, duplicating biases ~\cite{lee-etal-2022-deduplicating}, and social biases ~\cite{narayanan-venkit-etal-2023-nationality}.
    \item Another disadvantage of incorporating LLMs commercially is the \textbf{expensive cost} \cite{10.1145/3381831}. For an average business entity, applying LLMs for business use is barely feasible. It is almost impossible for a non-tech company to customize and train a GPT model of its own because they don't have the resources and talents to conduct such a big project ~\cite{musser2023cost}, while frequent API calls to third-party LLM providers like OpenAI can be extremely expensive, not to mention there are a very limited number of such providers in certain areas. 
    \item The \textbf{oblivion} problem of LLMs has been controversial because LLMs are found to tend to forget information from previous inputs. It is studied that LLMs also exhibit the behavior of catastrophic forgetting ~\cite{luo2023empirical} as neural networks do ~\cite{kemker2017measuring}.
\end{enumerate*}

While generative models like LLMs use vector data embeddings to represent unstructured data in our real world, users need a reliable data system to manage and retrieve a considerable amount of data with cost-efficiency. Yet another increasingly popular field that seems orthogonal to LLMs is purpose-built databases for AI that support vector data storage and its efficient retrieval at scale, also known as VecDB. With efficient combinations of VecDBs and LLMs, VecDBs can either be incorporated as an external knowledge base with efficient retrieval providing domain-specific knowledge for LLMs, a memory for LLMs saving previous related chat contents for each user's dialog tab, or a semantic cache, the aforementioned problems of LLMs can be seamlessly solved. 

In the light of lacking papers that introduce LLMs in the view of VecDB, this survey aims to picture how VecDBs can be potential solutions to refine LLMs' known shortcomings in previous works and hopes to offer a unique perspective of future directions in the intersection that is fertile of research opportunities. 

\begin{table*}[ht]
\centering
\begin{tabular}{@{}lllllll@{}}
\hline \hline
\thead{Name} & \multicolumn{2}{c}{\thead{Supported Data}} & \multicolumn{2}{c}{\thead{Supported Query}} & \multicolumn{2}{c}{\thead{Vector Index}} \\
\cmidrule(lr){2-3} \cmidrule(lr){4-5} \cmidrule(lr){6-7}
\thead{(Version with year)}& \thead{Vec. Dim.} & \thead{Type} & \thead{Filter} & \thead{Multi-Vec.} & \thead{Graph} & \thead{IVF} \\
\midrule
ChromaDB (2022) & 1536 & Vec. & $\checkmark$ & & $\checkmark$ & \\
Manu (2022) & 32768 & Vec.+ Ftx. & $\checkmark$ & $\checkmark$ & $\checkmark$ & $\checkmark$\\
Milvus (2021) & 32768 & Vec.+ Ftx. & $\checkmark$ & $\checkmark$ & $\checkmark$ & $\checkmark$ \\
Pinecone (2019) & 20000 & Vec. & $\checkmark$ & $\checkmark$ & $\checkmark$ & $\checkmark$\\
Weaviate (2019) & 65535 & Vec.+ Ftx. & $\checkmark$ & $\checkmark$ & $\checkmark$ & \\
Qdrant (2021) & 4096 & Vec.+ Ftx. & $\checkmark$ & $\checkmark$ & $\checkmark$ & \\
\midrule
Amazon OpenSearch (v2.9, 2023)& 16,000 & Ftx. & $\checkmark$ & $\checkmark$ & $\checkmark$ & $\checkmark$ \\
Elastic Search (v8.0, 2022) & 4096 & Ftx. & $\checkmark$ & $\checkmark$ & $\checkmark$ & $\checkmark$  \\
\midrule
AnalyticDB-V (2020) & $\ge$512 & Rel.+Ftx. & $\checkmark$ & & $\checkmark$ & $\checkmark$ \\
PostgreSQL-pgvector (2021) & 2000 & Rel.+ Ftx. & $\checkmark$ & $\checkmark$ & $\checkmark$ & $\checkmark$ \\
MongoDB Atlas (v6.0, 2023) & 2048 & NoSQL+ Ftx. & $\checkmark$ & & $\checkmark$ & \\
MyScale (2023) & 1536 & Rel. & $\checkmark$ & & $\checkmark$ & \\
\hline \hline
\end{tabular}
\caption{Comparison of mainstream all-level databases supporting vector data (the version number represents the first version that supports vector search and its release date). Abbreviations note: \textit{Ftx.}, \textit{Rel.}, \textit{Vec.} stand for full-text search, relational database, and vector search.}
\label{tab:vecDBs}
\end{table*}

\section{Background} \label{sec:bg}

\subsection{Large Language Models} \label{sec:llm}

Over the last half-decade, we have witnessed the groundbreaking success of LLMs, marking a significant milestone in the field of NLP. LLMs have revolutionized our views on the approaches to understanding and generating human languages by machines. 

\subsubsection{Development of Language Models}
With the advent of neural networks, the field of NLP has undergone a transformative shift, which began with the introduction of Recurrent Neural Networks (RNNs) ~\cite{zaremba2014recurrent}. RNNs provide a way to process sequences of words and capture temporal dependencies in text. And 2 of its famous variants, Gated Recurrent Units (GRUs) ~\cite{chung2014empirical} and Long Short-Term Memory networks (LSTMs) ~\cite{shi2015convolutional} address the limitations of RNNs in handling problems of vanishing and exploding gradients. 

The next pivotal milestone for language models is the widespread adoption of the Transformer architecture ~\cite{vaswani2017attention} which has set a new standard for language models. The multi-head self-attention modules and cross-attention modules in the encoders and decoders enable the model to capture long-range dependencies, parallel processing, and contextual understanding. Furthermore, the model starts to rapidly grow in size in the Transformer era. More importantly, the Transformer represents a paradigm shift in the NLP field, and the success of the Transformer architecture becomes the significant backbone of LLMs.

\subsubsection{Power of Pre-training}

The scaling laws proposed by OpenAI ~\cite{kaplan2020scaling} highlight a critical trend: the scaling of Pretrained Language Models (PLMs), in terms of both model size and data volume, leads to significant improvement in downstream tasks. This is evidenced by the development of increasingly larger PLMs, such as the 1.76-trillion-parameter GPT-4 ~\cite{achiam2023gpt} and the 540-billion-parameter PaLM ~\cite{chowdhery2023palm}. Unlike their smaller predecessors, like the 330-million-parameter BERT ~\cite{devlin2018bert} and 1.5-billion-parameter GPT-2 ~\cite{radford2019language}, these large-scale models demonstrate unique behaviors and emergent abilities ~\cite{wei2022emergent} in various tasks like zero-shot learning, which is extremely challenging for the small-scale models.

The power of pre-training lies in its ability to provide models with a general understanding of the languages, which is then tailored through additional training, known as fine-tuning, for specific tasks. The advantage is that the model doesn't start from scratch when learning a new task; it builds upon a rich, pre-existing foundation of language understanding. This enables what is known as transfer learning ~\cite{yosinski2014transferable}. 

\subsubsection{Are We There Yet? Challenges Faced by Pure LLMs}

Although LLMs have achieved remarkable success, they face significant challenges that cannot be overlooked. Three primary challenges are particularly noteworthy in the context of pure LLMs. Addressing these challenges is crucial for the continued advancement and practical application of these models. We have outlined these challenges as follows:

\begin{itemize}
    \item \textbf{Model Editing} involves integrating more flexible architectures into models, enabling targeted updates or edits without the need for complete retraining. This presents a complex challenge: alterations in one part of the model can unexpectedly affect other areas. Ensuring that these updates neither degrade performance nor introduce biases is a difficult task. Model editing also includes elements such as personality, emotions, opinions, and beliefs. While these aspects have received some attention, they largely remain unexplored territories ~\cite{yao2023editing}. Ethical concerns, including data biases ~\cite{raffel2020exploring}, privacy issues ~\cite{li2023privacy}, and the potential misuse of generative content ~\cite{ganguli2022red}, are significant challenges in model editing.

    \item \textbf{Hallucinations} in LLMs ~\cite{ji2023survey} pose a major challenge, undermining the reliability of their outputs. LLMs often generate plausible yet factually incorrect or nonsensical information ~\cite{huang2023survey}. Several factors contribute to this issue. First, LLMs lack domain-specific knowledge, primarily due to their training on public datasets ~\cite{penedo2023refinedweb}, limiting their ability to answer domain-specific questions. Additionally, updating LLMs with real-time knowledge is a challenge, leading to outdated internal knowledge in a rapidly changing world ~\cite{onoe-etal-2022-entity}. 

    \item \textbf{Substantial computational resources} required for training and operating LLMs raise environmental and accessibility concerns ~\cite{strubell2019energy}, contradicting the principles of sustainable AI ~\cite{WuRaghavendraGupta2022}. The escalating sizes of LLMs and their training datasets significantly influence training and inference costs. The environmental footprint of these technologies is substantial, with significant greenhouse gas emissions and energy consumption during model training and tuning ~\cite{Kaack2022}. This situation necessitates a critical evaluation of AI's environmental impact, encouraging the development of more energy-efficient models and the adoption of green computing practices ~\cite{vanWynsberghe2021}.

\end{itemize}

Transitioning to VecDBs could potentially comprehensively address these challenges. VecDBs, with their efficient data structuring and retrieval capabilities, offer a promising solution for enhancing model editing flexibility, reducing hallucinations by improving domain-specific responses, and optimizing computational resource utilization, thereby aligning more closely with the principles of sustainable AI.

\subsection{Vector Databases, the V-factor} \label{sec:vecdb}
\subsubsection{A distinct database optimized for vector data}
While LLMs like ChatGPT are relatively new concepts, database management systems (DBMS) have been thoroughly developed and applied in many aspects in the last 60 years of history and are well-recognized for their consistent stability and universality for structured data with fixed formats that excel well with computer storage. However, the development and wide application of deep learning models such as convolutional neural networks~\cite{he2016deep} and transformers~\cite{devlin2018bert}, enable the embedding of unstructured multi-modal data like images and text mapped into corresponding fixed-length vector representations, which include the \textbf{high-dimensional semantic features} of original data, and the semantic similarities are naturally represented by distances between vectors, requiring a new type of DBMS that is specifically designed for handling vector data operations, especially search and storage.

There are many kinds of databases, whereas VecDBs are the only category of databases that naturally support diverse unstructured data with efficient storage, indexing, and retrieval. As a result of various blooming applications on the cloud, various data sources exist in many different formats and diverse places. Unlike traditional databases that require structured or semi-structured data that must convey a few restrictions and formats, VecDBs are purpose-designed for storing the deep learning embedding of various unstructured data that has emerged in real-world applications. On the other hand, distinct from traditional DBMSs that search for exact values within databases, the semantic feature of vectors in VecDBs needs an approximate search for vectors, which searches approximate top-k nearest-distance neighbors within the high-dimensional base vector data space that do not necessarily require exact matches, representing top-k semantically close data.

\subsubsection{Efficient vector retrieval within VDBMS}
With the unstructured base data embedded into vectors with high dimensionality, calculating the k-nearest neighbors of a given query vector data can be expensive since it requires distance computation to every point in the dataset and maintaining the top-k results. Such an operation would result in a time complexity of $O(dN+N\log{k})$, where $d$ is the dimensionality and $N$ is the number of vectors, for exhaustively searching top-k results using pair-wise distance calculation and a heap to keep top-k results.

Since brute-force search is time-consuming and computationally expensive, this calls for a more efficient search and storage technique with satisfying accuracy. Vector indexing can solve both of these problems, which optimized for ANN search within VDBMS include tree-based~\cite{flann09,tao2009quality}, hash-based~\cite{lsh08}, Product Quantization (PQ)~\cite{JegouDS11}, and graph-based methods~\cite{hnsw18}, while the ANN Benchmarks~\footnote{\url{https://ann-benchmarks.com/}} has also showcased the great performance gap between brute-force search and index-enhanced ANN search. All these operations could be efficiently done based on the vector indexes, which are optimally designed collections of vectors deployed together for approximate nearest neighbor (ANN) search. The ANN Benchmarks have also showcased the great performance gap between brute-force search and index-enhanced ANN search. Also,  according to them, among all these indexes, the graph-based Hierarchical Navigable Small Worlds (HNSW) provides state-of-the-art performance and capability with great universality and is widely used within most VecDB management systems, including the examples we mentioned in Table~\ref{tab:vecDBs}. 

\subsubsection{A feasible solution of knowledge base for LLMs}
%% Why can we bridge vector db to LLMs? 
Traditional full-text and keyword search engines, such as Elastic Search and Amazon OpenSearch, are based on term frequency metrics like BM25, which has proved practical for many search applications. However, such search techniques require significant investment in time and expertise to tune them to account for the meaning or relevance of the terms searched. On the other hand, they lack multi-modal storage and retrieval support, as VecDBs do.

Nevertheless, VecDBs are capable of solving the aforementioned problem by simply transforming data into vectors and leveraging their retrieval efficiency; thus, the emergence of VecDBs (as shown in Table~\ref{tab:vecDBs}) has greatly influenced machine learning systems and their multi-modal applications. An intuitive application using vector search is searching images with an image on search engines built on VecDBs~\cite{wang2021milvus} like Google Image Search~\footnote{\url{https://cloud.google.com/vertex-ai/docs/vector-search/overview}}, which uses one input image to find similar images on the internet. While VecDBs boast their unique search capability and efficiency for unstructured data, they are just a back-end factor for manifold applications. To maximize their abilities, this leads us to an interesting question: since LLMs use embeddings to represent text as vectors, can developers combine VecDBs and LLMs to overcome the aforementioned challenges that are inherited in LLM applications? The answer is yes.

%  What can vector database bring to LLMs, and advantages
As a kind of database encapsulated with vector search in vector data that represents real-world information within high dimensionalities, VecDBs are well-capable for retrieval applications~\cite{retrieval-lm-tutorial} incorporating LLMs because LLM applications are generally read-intensive, not requiring many write-related changes, especially data deletes~\cite{pan2023survey}. On the other hand, VecDBs can efficiently manage and warehouse vector data required and generated by LLMs, thus providing a solid data cornerstone for both LLMs and their applications. While LLMs are often limited by domain knowledge that cannot be uploaded or distributed due to security and privacy concerns, domain knowledge is often unstructured data that can easily be embedded into VecDBs for efficient local retrieval and further integration with generative AIs. Moreover, the computing and storage resources of VecDBs are way cheaper than LLMs, since they do require costly GPUs and TPUs, thus achieving a cost-effective way of fast retrieval and durable (non-volatile) storage.

\section{Synergizing LLMs with VecDBs: Enhancements and Innovations} \label{sec:intersection}
\begin{figure*}
    \centering
    \includegraphics[width=\textwidth]{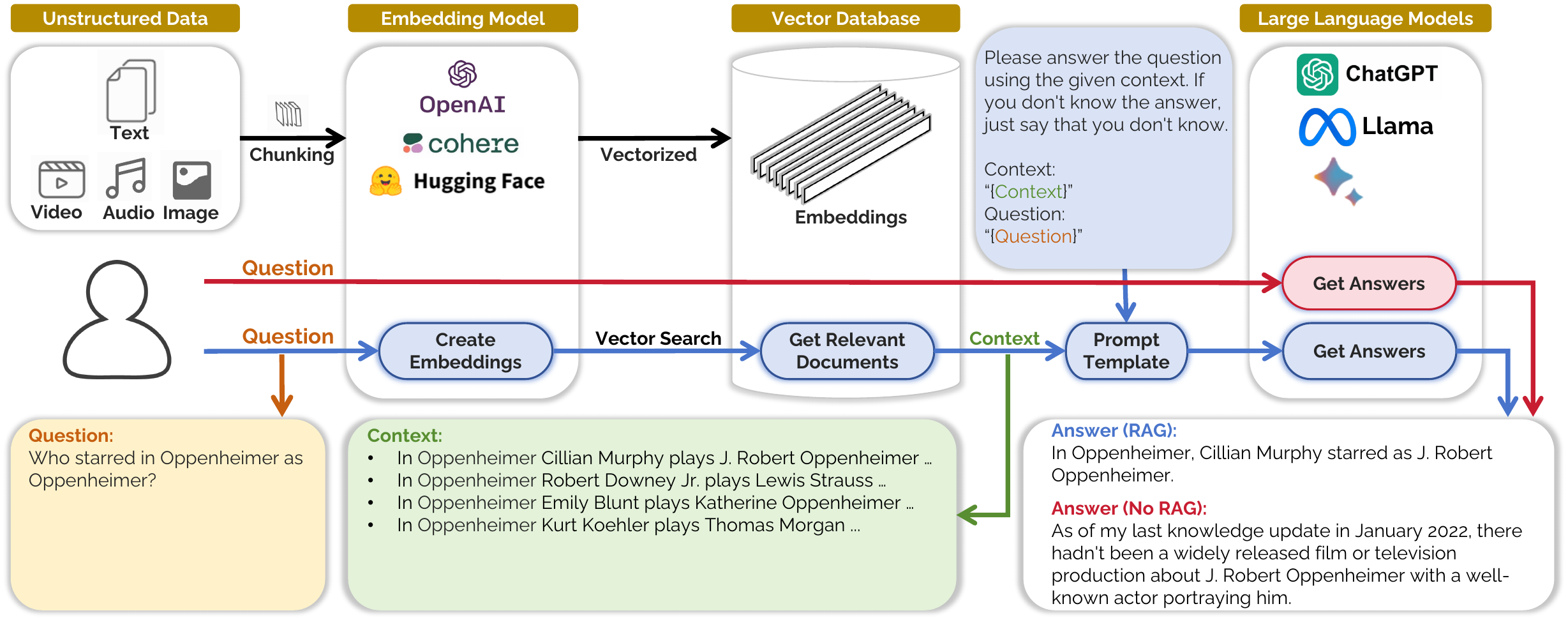}
    \caption{{A sample RAG framework that uses VecDBs.} }
    \label{fig:RAG-VecDB}
\end{figure*}

With recent rapid development in both areas, applications that utilize the combination of the two are also growing rapidly, and the most fruitful one is Retrieval-Augmented Generation (RAG), where VecDBs play a crucial role in enabling cost-effective data storage and efficient retrieval.

\subsection{VecDB as an External Knowledge Base: Retrieval-Augmented Generation (RAG)} \label{sec:rag}
\subsubsection{Development of RAG Paradigm}

As the release of ChatGPT casts a spotlight on LLMs to the public world, there is an increasing need to use AI chatbots as query and retrieval agents. But simply loading users' private data as input to LLMs is found to be incompetent in real-world applications. LLMs have been constrained by their limited token counts and the high costs associated with training and fine-tuning for every alteration of data, especially when dealing with personalized or business-specific responses that require real-time data updates. To address such concerns and needs, retrieval-augmented generation (RAG) emerges as a novel solution that addresses the challenges faced by LLMs in integrating and processing large and dynamic data in external databases, where VecDBs offer a solution by acting as external memory for LLMs. They allow for the segmentation of private data by converting them into vectors and storing these vectors efficiently for a quick retrieval process. Integrating with VecDBs enables LLMs to access and synthesize enormous amounts of data without the need for constant re-training, thereby overcoming their inherent limitations. 

The concept of RAG is devised to be a paradigm, and a common workflow of RAG is illustrated in Figure \ref{fig:RAG-VecDB}. There are essentially three main parts to a full run of the system: data storage, retrieval, and generation.

Data storage means establishing reliable external memory like VecDBs, and there are detailed recipes for this process \cite{han2023comprehensive}. It starts with data preprocessing, during which the original data is collected, cleaned, integrated, transformed, normalized, and standardized. The processed data is chunked into smaller segments because of the contextual limitations of LLMs. These segments of data are then converted by an embedding model into vectors, the semantic representations of the data, which are stored in VecDBs and will be used for vector search in later steps. A well-developed VecDB will properly index the data and optimize the retrieval process. The retrieval part starts with a user asking a question in the form of prompts to the same embedding model, which has generated the vector representation of the stored data and gained the vector embeddings of the question. The next step of the process is vector searching and scoring inside the VecDBs, which essentially involves computing similarity scores among vectors, and the database then identifies and retrieves the data segments that have the highest similarity scores (top $K$ in most RAG systems) compared to the query vector. These retrieved segments are then converted back from their vector format to their original form, which is typically the text of documents. In the generation part, LLMs are involved in generating the final answers. The retrieved documents, along with the user's question, are incorporated into a specifically chosen and designed prompt template. This template selection is based on the task type and aims to effectively merge the context with the question to form a coherent prompt. The selected LLM is provided with the prompt and generates the final answer. 

\subsubsection{RAG with VecDBs} 
In the prototype of the RAG paradigm, we use such an idea to overcome the hallucination problems of LLMs by providing domain-specific data with accurate instructions. where VecDBs serve as an external knowledge base to warehouse domain-specific data, then LLMs can easily handle massive data owned by users. 

Even though LLMs such as ChatGPT now have user-specified GPTs for specific uses, with just prompt engineering, their knowledge bases are still limited to the training data provided by OpenAI. However, by using VecDBs, users can pre-filter whatever they want it to look at, whereas that is difficult with prompt-engineering GPT assistants.

Many research and industry examples justified RAG's profound effect. Azure AI Search (formerly Cognitive Search) developed its vector search using Qdrant~\footnote{\url{https://learn.microsoft.com/en-us/azure/search/vector-search-overview}}. Another example is Pinecone's Canopy~\footnote{\url{https://www.pinecone.io/blog/canopy-rag-framework/}}, an open-source framework that leverages Pinecone's VecDB to build and host production-ready RAG systems. Companies like Spotify and Yahoo have adopted Vespa, an AI-driven online VecDB, and Yahoo uses it to help users directly chat with their mailbox, ask questions about their mailbox, and tell it to do things for them. This uses personalized data integrated with LLMs to form a RAG system~\footnote{\url{https://vespa.ai/}}. These examples demonstrate the strength of RAG combined with VecDBs to address the unique challenges faced by business enterprises in managing and extracting value from diverse datasets but also showcase how the aforementioned difficulties faced by LLMs can be solved via integration with VecDBs.

\subsection{VecDB as a Cost-effective Semantic Cache}
\label{sec:vecdb_cache}
The combination of VecDBs and LLM not only facilitates the profound application of LLM with RAG but also provides a new frontier for cost-effective end-to-end LLM applications.

\paragraph{Outrageous API Costs} LLM-based chatbots and agent systems heavily rely on LLM's output from API vendors; repeated or similar inquiries may lead to outrageous API costs.
\paragraph{API Bandwidth Limitations} Such chatbots and agent systems could also experience a bursty inquiry workload that may drown the system's bandwidth with explosive API calls coming within seconds, leading to the system's outage and reconfiguration.

\begin{figure}
    \centering
    \includegraphics[width = 0.9\linewidth]{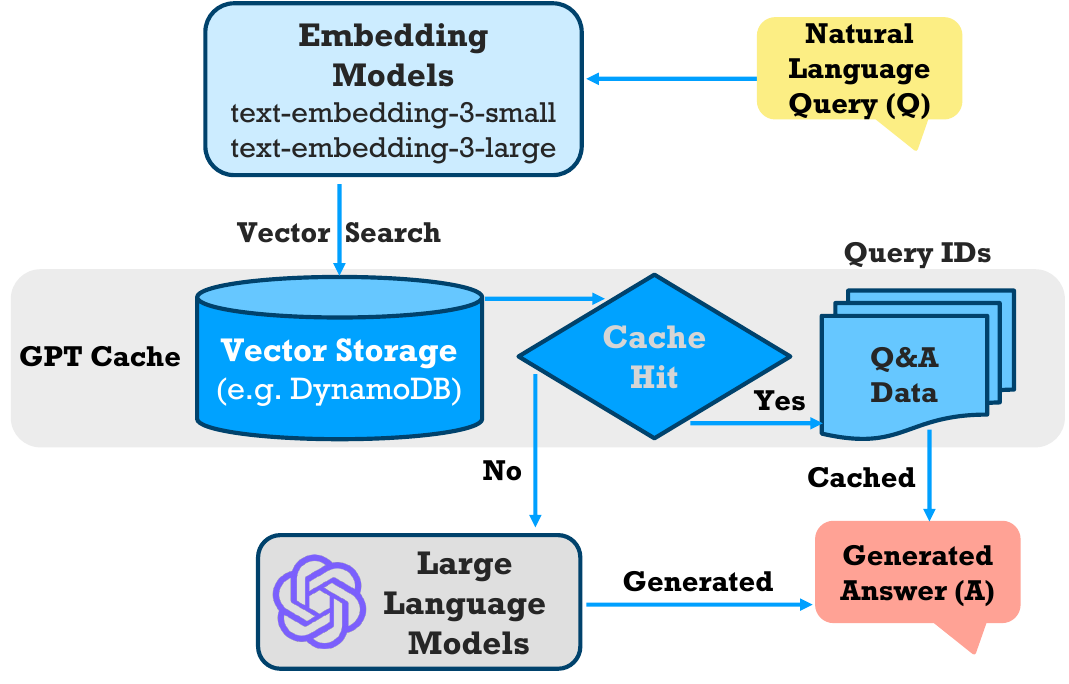}
    \caption{An overview of semantic cache for GPTs that utilizes VecDBs.}
    \label{fig:GPTCache}
\end{figure}

One of the primary benefits of integrating VecDBs with LLMs is the significant reduction in data operational costs~\cite{sanca2023scan}. GPT-Cache~\cite{bang2023gptcache}, a VecDB that serves as a semantic cache, as shown in Figure~\ref{fig:GPTCache}, for instance, stores responses to previously asked queries and works as a cache before calling LLM APIs. This caching mechanism means that the system doesn’t need to have API calls to wait for generated responses from scratch every time, reducing the number of costly API calls to the LLM. Moreover, this approach also speeds up the response time, enhancing the user experience.

VecDBs enhance the LLMs' ability to retrieve and utilize relevant information by indexing vast amounts of previous Q\&A data and mapping them into a vector space. Instead of caches in computer systems that require exact hash-match, this allows more precise semantic matching of queries with existing knowledge and results in responses that are not only generated based on the LLM's training data but also informed by the most relevant and recent information available in the VecDBs. 

\subsection{VecDB as a Reliable Memory of GPTs} 
 \label{sec:vecdb_memo_of_gpt}

\paragraph{Oblivion} When using LLM Q\&A applications like ChatGPT, LLMs are likely to completely forget the content and information of your previous conversation, even within the same chat tab. 

VecDBs serve as a robust memory layer~\cite{10337079} that addresses one of the intrinsic limitations of LLMs: the static nature of their knowledge. While LLMs excel in generating human-like text based on patterns learned during training, they cannot inherently update their knowledge base dynamically and, thus, may lack the few-shot learning ability. VecDBs bridge this gap by offering a storage solution that can be continually updated with new information, ensuring that the LLM's responses are informed by the most current and relevant data available.

The VecDB and LLM combination brings forth a synergy where the LLM provides context and understanding for user queries, while the VecDB offers a precise mechanism for storing and retrieving the relevant vectors. This collaborative approach allows for more accurate, relevant, and efficient responses to complex queries, which would be challenging for either system to address independently.

A VecDB integrated with an LLM facilitates real-time learning and adaptation. As new data is ingested into the VecDB, the LLM can immediately leverage this updated repository to refine its responses. This capability is pivotal for applications requiring up-to-the-minute accuracy, such as financial analysis, news dissemination, and personalized recommendations.

\section{Expanding Horizons: RAG Advancements}
\subsection{Multimodality of RAG} 
RAG has now evolved to handle a wide range of data types by lending the power of multimodal models. The impressive achievements of LLMs have inspired significant advancements in vision-language research. DALL-E from OpenAI introduced a Transformer-based approach for converting text to images, treating images as sequences of discrete tokens. Subsequent improvements in the text-to-image area ~\cite{zhang2023texttoimage} have been achieved through methods like model scaling, pre-training, and enhanced image quantization models. BLIP-2 ~\cite{li2023blip2} uses static image encoders with LLMs for efficient visual language pre-training, facilitating direct image-to-text transformations. Flamingo ~\cite{alayrac2022flamingo} presented a visual language model for text generation, showcasing remarkable adaptability and leading performance across various vision-and-language tasks. CM3 ~\cite{aghajanyan2022cm3} trained a randomly masked model on a large HTML corpus and showed that the model is capable of generating images and text. FROMAGe ~\cite{koh2023grounding} gains robust multimodal capabilities for few-shot learning solely from image-caption pairs, unlike other models that necessitate large-scale, interwoven image-text data from the websites.

To import speech data to RAG systems, Wav2Seq ~\cite{wu2022wav2seq} allows for efficient pre-training without the need for transcriptions, using techniques like k-means clustering and byte-pair encoding to generate pseudo-subwords from speech. The Large Language and Speech Model (LLaSM) ~\cite{shu2023llasm} is an end-to-end trained, large multi-modal speech-language model equipped with cross-modal conversational skills and proficient in understanding and responding to combined speech-and-language directives. Videos are also made available for certain types of RAG systems. Vid2Seq ~\cite{yang2023vid2seq} enhances language models with specific temporal indicators for predicting event limits and textual descriptions in a single output sequence. 

\subsection{Retrieval Optimizations of RAG} To better harness the knowledge from various types of data, kNN-LMs ~\cite{khandelwal2020generalization} explore how incorporating nearest neighbor search into language models can enhance their ability to generalize by effectively leveraging memorized examples. EASE ~\cite{nishikawa-etal-2022-ease} is distinctive in its use of entities as a strong indicator of text semantics, providing rich training signals for sentence embedding. In-context RALM ~\cite{ram-etal-2023-context} proves that with the language model architecture unchanged, simply appending grounding documents to the input will improve the performance. SURGE ~\cite{kang2023knowledge} enhances dialogue generation by incorporating context-relevant sub-graphs from a knowledge graph. Another work that combines knowledge graphs and LLMs is RET-LLM ~\cite{modarressi2023retllm}, which is designed to equip LLMs with a general write-read memory unit. These studies have focused on retrieval granularity and data structuring levels, with coarse granularity providing more, but less precise, information. Conversely, structured text retrieval offers detailed information at the cost of efficiency.

To utilize both internal knowledge and external resources, SKR ~\cite{yu2023generate} improves LLMs' ability by enabling them to assess what they know and identify when to seek external information to answer questions more effectively. Selfmem ~\cite{cheng2023lift} enhances retrieval-augmented text generation by creating an unbounded memory pool using the model's output and selecting the best output as memory for subsequent rounds. FLARE ~\cite{jiang2023active} uses predictions of upcoming sentences to anticipate future content and retrieve relevant documents for regenerating sentences, especially when they contain low-confidence tokens. Atlas ~\cite{izacard2022atlas} demonstrates impressive performance in tasks like question-answering and fact-checking, outperforming much larger models despite having fewer parameters. Many other works like these also aim to make the RAG system more efficient and competent.

\section{Discussion: Challenges and Future Work} \label{sec:discussion}
\subsubsection{Are vector searches all you need?}  
Although VecDBs offer a cost-efficient modality for information retrieval within LLMs' frameworks, their utility in traditional relational database operations remains limited. Specifically, vector-based search methodologies are still not well optimized for operations such as post-query filtering, comprehensive full-text searches, and keyword search mechanisms that are fundamental to conventional database systems and search engines. In the context of diverse applications for different users with multiple needs, vector search, or any single search approach, is not a one-size-fits-all solution. For example, for searching exact transaction data, users may have queries of very specific keywords to look up in a huge pool of data; giving them semantic approximate results totally would not make sense, resulting in a one-step forward yet two-step back. On the other hand, traditional DBMSs are well-developed for efficiently handling structured data within their transactional and analytical applications, while dedicated VecDBs are not yet.

This distinction underscores a potential gap in the functional alignment of VecDBs with previously well-established data retrieval paradigms, necessitating the development of hybrid search algorithms that can seamlessly integrate vector search with traditional relational data engine capabilities.

\subsubsection{VDBMS with multi-modality}
In real-world data representation and search, we may look for multi-modal data entries that are not restricted to a specific kind of structured, semi-structured, or unstructured data, but with any random combination of them, e.g., users can look for a knowledge graph from an interacted prompt of image and description. While searches using uni-modal data may lack the prospect of providing informative and more contextually appropriate search results, multi-modal data search and its hybrid processing also present a non-trivial challenge for VecDBs' storage and retrieval~\cite{wang2023must}, since integrating and merging multi-modal data require efficient preprocessing with multi-modal encoders, multi-modal storage indexes, but also ranking weight assignment and multi-modal data fusion. 

\subsubsection{Data preprocessing}
While \textbf{text embedding} is considered efficient for processing long text and its retrieval, it comes with a dazzling challenge for building every text-based knowledge database. It is observed that vector retrieval would have fundamentally better performance when precise chunks of text with clear meaning are applied. This raises an interesting question: how do we have a proper (probably unified) embedding methodology for every raw text?

On the other hand, \textbf{dimension reduction} for vector data is considered important since compressing high-dimensional vectors will use less memory and disk resources. From this view, the current vector data dimension is not cost-effective, as the vectors are large, e.g., a 1536-dimensional vector embedded only by OpenAI's lightweight model text-embedding-3-small~\footnote{\url{https://platform.openai.com/docs/guides/embeddings}} is about 6 kilobytes; with scaling up to 1 billion, the data size would be 6 terabytes. Also with \textit{the curse of dimensionality} theory, data sparsity increases with dimensionality, resulting in hardship in distinguishing meaningful similarities.  Moreover, current dimension reduction algorithms like PQ contribute to a memory-efficient vector data representation, whereas lossy data compression still hurts the recall quality~\cite{ChenW21}. This requires a more efficient vector data dimension reduction algorithm for VecDBs and searches.

\subsubsection{Data management systems for LLMs with multi-tenancy}
Both for VecDBs and LLMs, this challenge is particularly in preventing data leaks between tenants and maintaining the isolation and security of distinct tenants' data while ensuring efficiency, where isolation is crucial for privacy and security. On the other hand, it must be balanced with the need for dynamic resource sharing to ensure resources are flexibly allocated and deallocated.

\subsubsection{Cost-effective and scalable data storage and retrieval}
While state-of-the-art vector search algorithms are mainly memory-based, memory resources in the cloud are considered expensive and volatile. With billion-scale amounts of data, naively handling them in memory resources is bankrupting. On the other hand, disk-based search methods have downgraded performance that does not satisfy real-time applications with low latency tolerance. This calls for a more cost-effective way to ensure the performance of LLM applications.

\subsubsection{Knowledge conflict}
Conflicting knowledge from the same or different knowledge bases presents a non-trivial challenge for both humans and LLMs to distinguish the correct piece of knowledge. This conflict becomes particularly prominent when integrating multiple knowledge bases where each potentially carries its own biases or inaccuracies, such as when different datasets provide contradictory information about the same fact or when biases inherent in data sources lead to inconsistent conclusions. While VecDBs cannot naturally distinguish, resolving such conflicts requires robust conflict resolution strategies that can assess the reliability of sources and the context of data to determine the most accurate information.

\section{Conclusion}
In this paper, we present a systematic review of recent advances in combinations of LLMs and VecDBs. We summarize the rationales for viewing LLMs with VecDBs and also introduce certain applications that combine LLMs and VecDBs with distinct prototypes that categorize existing works from various perspectives and interdisciplinary studies. Our study also demonstrates both the research and engineering challenges in this fast-growing field and suggests directions for future work.

\bibliographystyle{named}

\begin{thebibliography}{}

\bibitem[\protect\citeauthoryear{Achiam \bgroup \em et al.\egroup }{2023}]{achiam2023gpt}
Josh Achiam, Steven Adler, et~al.
\newblock Gpt-4 technical report.
\newblock {\em arXiv preprint arXiv:2303.08774}, 2023.

\bibitem[\protect\citeauthoryear{Aghajanyan \bgroup \em et al.\egroup }{2022}]{aghajanyan2022cm3}
Armen Aghajanyan, Bernie Huang, et~al.
\newblock Cm3: A causal masked multimodal model of the internet.
\newblock {\em arXiv preprint arXiv:2201.07520}, 2022.

\bibitem[\protect\citeauthoryear{Alayrac \bgroup \em et al.\egroup }{2022}]{alayrac2022flamingo}
Jean Alayrac, Jeff Donahue, et~al.
\newblock Flamingo: a visual language model for few-shot learning.
\newblock {\em arXiv preprint arXiv:2204.14198}, 2022.

\bibitem[\protect\citeauthoryear{Andoni and Indyk}{2008}]{lsh08}
Alexandr Andoni and Piotr Indyk.
\newblock Near-optimal hashing algorithms for approximate nearest neighbor in high dimensions.
\newblock {\em Comm. ACM}, 51(1):117--122, 2008.

\bibitem[\protect\citeauthoryear{Asai \bgroup \em et al.\egroup }{2023}]{retrieval-lm-tutorial}
Akari Asai, Sewon Min, et~al.
\newblock Retrieval-based language models and applications.
\newblock In {\em ACL}, pages 41--46, 2023.

\bibitem[\protect\citeauthoryear{Bang}{2023}]{bang2023gptcache}
Fu~Bang.
\newblock Gptcache: An open-source semantic cache for llm applications enabling faster answers and cost savings.
\newblock In {\em NLP-OSS Workshop}, pages 212--218, 2023.

\bibitem[\protect\citeauthoryear{Bender \bgroup \em et al.\egroup }{2021}]{10.1145/3442188.3445922}
Emily~M Bender, Timnit Gebru, et~al.
\newblock On the dangers of stochastic parrots: Can language models be too big?
\newblock In {\em FAccT}, page 610–623, 2021.

\bibitem[\protect\citeauthoryear{Brown \bgroup \em et al.\egroup }{2020}]{brown2020language}
Tom Brown, Benjamin Mann, et~al.
\newblock Language models are few-shot learners.
\newblock In {\em NeurIPS}, pages 1877--1901, 2020.

\bibitem[\protect\citeauthoryear{Chen \bgroup \em et al.\egroup }{2021}]{ChenW21}
Qi~Chen, Bing Zhao, et~al.
\newblock Spann: Highly-efficient billion-scale approximate nearest neighbor search.
\newblock In {\em NeurIPS 2021}, 2021.

\bibitem[\protect\citeauthoryear{Chowdhery \bgroup \em et al.\egroup }{2023}]{chowdhery2023palm}
Aakanksha Chowdhery, Sharan Narang, et~al.
\newblock Palm: Scaling language modeling with pathways.
\newblock {\em JMLR}, 24(240):1--113, 2023.

\bibitem[\protect\citeauthoryear{Chung \bgroup \em et al.\egroup }{2014}]{chung2014empirical}
Junyoung Chung, Caglar Gulcehre, et~al.
\newblock Empirical evaluation of gated recurrent neural networks on sequence modeling.
\newblock {\em arXiv preprint arXiv:1412.3555}, 2014.

\bibitem[\protect\citeauthoryear{Devlin \bgroup \em et al.\egroup }{2018}]{devlin2018bert}
Jacob Devlin, Ming-Wei Chang, et~al.
\newblock Bert: Pre-training of deep bidirectional transformers for language understanding.
\newblock {\em arXiv preprint arXiv:1810.04805}, 2018.

\bibitem[\protect\citeauthoryear{Ganguli \bgroup \em et al.\egroup }{2022}]{ganguli2022red}
Deep Ganguli, Liane Lovitt, et~al.
\newblock Red teaming language models to reduce harms: Methods, scaling behaviors, and lessons learned.
\newblock {\em arXiv preprint arXiv:2209.07858}, 2022.

\bibitem[\protect\citeauthoryear{Han \bgroup \em et al.\egroup }{2023}]{han2023comprehensive}
Yikun Han, Chunjiang Liu, et~al.
\newblock A comprehensive survey on vector database: Storage and retrieval technique, challenge.
\newblock {\em arXiv preprint arXiv:2310.11703}, 2023.

\bibitem[\protect\citeauthoryear{He \bgroup \em et al.\egroup }{2016}]{he2016deep}
Kaiming He, Xiangyu Zhang, et~al.
\newblock Deep residual learning for image recognition.
\newblock In {\em CVPR}, pages 770--778, 2016.

\bibitem[\protect\citeauthoryear{Huang \bgroup \em et al.\egroup }{2023}]{huang2023survey}
Lei Huang, Weijiang Yu, et~al.
\newblock A survey on hallucination in large language models: Principles, taxonomy, challenges, and open questions.
\newblock {\em arXiv preprint arXiv:2311.05232}, 2023.

\bibitem[\protect\citeauthoryear{Izacard \bgroup \em et al.\egroup }{2022}]{izacard2022atlas}
Gautier Izacard, Lewis Patrick, et~al.
\newblock Atlas: Few-shot learning with retrieval augmented language models.
\newblock {\em arXiv preprint arXiv:2208.03299}, 2022.

\bibitem[\protect\citeauthoryear{J{\'{e}}gou \bgroup \em et al.\egroup }{2011}]{JegouDS11}
Herv{\'{e}} J{\'{e}}gou, Matthijs Douze, et~al.
\newblock {Product Quantization for Nearest Neighbor Search}.
\newblock {\em TPAMI}, 33(1):117--128, 2011.

\bibitem[\protect\citeauthoryear{Ji \bgroup \em et al.\egroup }{2023}]{ji2023survey}
Ziwei Ji, Nayeon Lee, et~al.
\newblock Survey of hallucination in natural language generation.
\newblock {\em ACM Computing Surveys}, 55(12):1--38, 2023.

\bibitem[\protect\citeauthoryear{Jiang \bgroup \em et al.\egroup }{2023}]{jiang2023active}
Zhengbao Jiang, Frank~F Xu, et~al.
\newblock Active retrieval augmented generation.
\newblock {\em arXiv preprint arXiv:2305.06983}, 2023.

\bibitem[\protect\citeauthoryear{Kaack \bgroup \em et al.\egroup }{2022}]{Kaack2022}
Lynn~H Kaack, Priya~L Donti, et~al.
\newblock Aligning artificial intelligence with climate change mitigation.
\newblock {\em Nature Climate Change}, 12:518--527, 2022.

\bibitem[\protect\citeauthoryear{Kang and Kwak}{2023}]{kang2023knowledge}
Minki Kang and Jin~M Kwak.
\newblock Knowledge graph-augmented language models for knowledge-grounded dialogue generation.
\newblock {\em arXiv preprint arXiv:2305.18846}, 2023.

\bibitem[\protect\citeauthoryear{Kaplan \bgroup \em et al.\egroup }{2020}]{kaplan2020scaling}
Jared Kaplan, Sam McCandlish, et~al.
\newblock Scaling laws for neural language models.
\newblock {\em arXiv preprint arXiv:2001.08361}, 2020.

\bibitem[\protect\citeauthoryear{Kemker \bgroup \em et al.\egroup }{2017}]{kemker2017measuring}
Ronald Kemker, Marc McClure, et~al.
\newblock Measuring catastrophic forgetting in neural networks.
\newblock {\em arXiv preprint arXiv:1708.0207}, 2017.

\bibitem[\protect\citeauthoryear{Khandelwal \bgroup \em et al.\egroup }{2020}]{khandelwal2020generalization}
Urvashi Khandelwal, Levy Omer, et~al.
\newblock Generalization through memorization: Nearest neighbor language models, 2020.

\bibitem[\protect\citeauthoryear{Koh \bgroup \em et al.\egroup }{2023}]{koh2023grounding}
Jing~Yu Koh, Ruslan Salakhutdinov, et~al.
\newblock Grounding language models to images for multimodal inputs and outputs.
\newblock In {\em ICML}, 2023.

\bibitem[\protect\citeauthoryear{Lee \bgroup \em et al.\egroup }{2022}]{lee-etal-2022-deduplicating}
Katherine Lee, Daphne Ippolito, et~al.
\newblock Deduplicating training data makes language models better.
\newblock In {\em ACL}, pages 8424--8445, 2022.

\bibitem[\protect\citeauthoryear{Li \bgroup \em et al.\egroup }{2023a}]{li2023privacy}
Haoran Li, Yulin Chen, et~al.
\newblock Privacy in large language models: Attacks, defenses and future directions.
\newblock {\em arXiv preprint arXiv:2310.10383}, 2023.

\bibitem[\protect\citeauthoryear{Li \bgroup \em et al.\egroup }{2023b}]{li2023blip2}
Junnan~Li Li, Dongxu Li, et~al.
\newblock Blip-2: Bootstrapping language-image pre-training with frozen image encoders and large language models.
\newblock {\em arXiv preprint arXiv:2301.12597}, 2023.

\bibitem[\protect\citeauthoryear{Lin \bgroup \em et al.\egroup }{2022}]{lin-etal-2022-truthfulqa}
Stephanie Lin, Jacob Hilton, and Owain Evans.
\newblock {T}ruthful{QA}: Measuring how models mimic human falsehoods.
\newblock In Smaranda Muresan, Preslav Nakov, et~al., editors, {\em ACL}, pages 3214--3252, 2022.

\bibitem[\protect\citeauthoryear{Luo \bgroup \em et al.\egroup }{2023}]{luo2023empirical}
Yun Luo, Zhen Yang, et~al.
\newblock An empirical study of catastrophic forgetting in large language models during continual fine-tuning.
\newblock {\em arXiv preprint arXiv:2308.08747}, 2023.

\bibitem[\protect\citeauthoryear{Malkov and Yashunin}{2018}]{hnsw18}
Yury~A Malkov and Dmitry~A Yashunin.
\newblock {Efficient and Robust Approximate Nearest Neighbor Search Using Hierarchical Navigable Small World Graphs}.
\newblock {\em PAMI}, 42(4):824--836, 2018.

\bibitem[\protect\citeauthoryear{Modarressi \bgroup \em et al.\egroup }{2023}]{modarressi2023retllm}
Ali Modarressi, Ayyoob Imani, et~al.
\newblock Ret-llm: Towards a general read-write memory for large language models.
\newblock {\em arXiv preprint arXiv:2305.14322}, 2023.

\bibitem[\protect\citeauthoryear{Muja and Lowe}{2009}]{flann09}
Marius Muja and David~G Lowe.
\newblock {Fast Approximate Nearest Neighbors with Automatic Algorithm Configuration}.
\newblock In {\em VISAPP}, pages 331--340, 2009.

\bibitem[\protect\citeauthoryear{Musser}{2023}]{musser2023cost}
Micah Musser.
\newblock A cost analysis of generative language models and influence operations.
\newblock {\em arXiv preprint arXiv:2308.03740}, 2023.

\bibitem[\protect\citeauthoryear{Narayanan~Venkit \bgroup \em et al.\egroup }{2023}]{narayanan-venkit-etal-2023-nationality}
Pranav Narayanan~Venkit, Sanjana Gautam, et~al.
\newblock Nationality bias in text generation.
\newblock In {\em EACL}, pages 116--122, 2023.

\bibitem[\protect\citeauthoryear{Nishikawa \bgroup \em et al.\egroup }{2022}]{nishikawa-etal-2022-ease}
Sosuke Nishikawa, Ryokan Ri, et~al.
\newblock {EASE}: Entity-aware contrastive learning of sentence embedding.
\newblock In {\em NAACL}, pages 3870--3885, 2022.

\bibitem[\protect\citeauthoryear{Onoe \bgroup \em et al.\egroup }{2022}]{onoe-etal-2022-entity}
Yasumasa Onoe, Michael Zhang, et~al.
\newblock Entity cloze by date: What {LM}s know about unseen entities.
\newblock In {\em NAACL Findings}, pages 693--702, 2022.

\bibitem[\protect\citeauthoryear{Pan \bgroup \em et al.\egroup }{2023}]{pan2023survey}
James~Jie Pan, Jianguo Wang, et~al.
\newblock Survey of vector database management systems.
\newblock {\em arXiv preprint arXiv:2310.14021}, 2023.

\bibitem[\protect\citeauthoryear{Penedo \bgroup \em et al.\egroup }{2023}]{penedo2023refinedweb}
Guilherme Penedo, Quentin Malartic, et~al.
\newblock The refinedweb dataset for falcon llm: Outperforming curated corpora with web data, and web data only.
\newblock {\em arXiv preprint arXiv:2306.01116}, 2023.

\bibitem[\protect\citeauthoryear{Radford \bgroup \em et al.\egroup }{2019}]{radford2019language}
Alec Radford, Jeffrey Wu, et~al.
\newblock Language models are unsupervised multitask learners.
\newblock {\em OpenAI Blog}, 1(8):9, 2019.

\bibitem[\protect\citeauthoryear{Raffel \bgroup \em et al.\egroup }{2020}]{raffel2020exploring}
Colin Raffel, Noam Shazeer, et~al.
\newblock Exploring the limits of transfer learning with a unified text-to-text transformer.
\newblock {\em JMLR}, 21(1):5485--5551, 2020.

\bibitem[\protect\citeauthoryear{Ram \bgroup \em et al.\egroup }{2023}]{ram-etal-2023-context}
Ori Ram, Yoav Levine, et~al.
\newblock In-context retrieval-augmented language models.
\newblock {\em TACL}, 11:1316--1331, 2023.

\bibitem[\protect\citeauthoryear{Sanca and Ailamaki}{2023}]{sanca2023scan}
Viktor Sanca and Anastasia Ailamaki.
\newblock E-scan: Consuming contextual data with model plugins.
\newblock In {\em VLDB Workshop}, 2023.

\bibitem[\protect\citeauthoryear{Schwartz \bgroup \em et al.\egroup }{2020}]{10.1145/3381831}
Roy Schwartz, Jesse Dodge, et~al.
\newblock Green ai.
\newblock {\em Commun. ACM}, 63(12):54–63, 2020.

\bibitem[\protect\citeauthoryear{Shi \bgroup \em et al.\egroup }{2015}]{shi2015convolutional}
Xingjian Shi, Zhourong Chen, et~al.
\newblock Convolutional lstm network: A machine learning approach for precipitation nowcasting.
\newblock In {\em NeurIPS}, 2015.

\bibitem[\protect\citeauthoryear{Shu \bgroup \em et al.\egroup }{2023}]{shu2023llasm}
Yu~Shu, Siwei Dong, et~al.
\newblock Llasm: Large language and speech model.
\newblock {\em arXiv preprint arXiv:2308.15930}, 2023.

\bibitem[\protect\citeauthoryear{Strubell \bgroup \em et al.\egroup }{2019}]{strubell2019energy}
Emma Strubell, Ananya Ganesh, et~al.
\newblock Energy and policy considerations for deep learning in nlp.
\newblock {\em arXiv preprint arXiv:1906.02243}, 2019.

\bibitem[\protect\citeauthoryear{Tao \bgroup \em et al.\egroup }{2009}]{tao2009quality}
Yufei Tao, Ke~Yi, et~al.
\newblock Quality and efficiency in high dimensional nearest neighbor search.
\newblock In {\em SIGMOD}, page 563–576, 2009.

\bibitem[\protect\citeauthoryear{Touvron \bgroup \em et al.\egroup }{2023}]{touvron2023llama}
Hugo Touvron, Thibaut Lavril, et~al.
\newblock Llama: Open and efficient foundation language models.
\newblock {\em arXiv preprint arXiv:2302.13971}, 2023.

\bibitem[\protect\citeauthoryear{Vaswani \bgroup \em et al.\egroup }{2017}]{vaswani2017attention}
Ashish Vaswani, Noam Shazeer, et~al.
\newblock Attention is all you need.
\newblock In {\em NeurIPS}, 2017.

\bibitem[\protect\citeauthoryear{Wang and et al.}{2023}]{wang2023must}
Mengzhao Wang and Xiangyu~Ke et~al.
\newblock Must: An effective and scalable framework for multimodal search of target modality.
\newblock {\em arXiv preprint arXiv:2312.06397}, 2023.

\bibitem[\protect\citeauthoryear{Wang \bgroup \em et al.\egroup }{2021}]{wang2021milvus}
Jianguo Wang, Xiaomeng Yi, et~al.
\newblock Milvus: A purpose-built vector data management system.
\newblock In {\em SIGMOD}, pages 2614--2627, 2021.

\bibitem[\protect\citeauthoryear{Wei \bgroup \em et al.\egroup }{2022}]{wei2022emergent}
Jason Wei, Yi~Tay, et~al.
\newblock Emergent abilities of large language models.
\newblock {\em arXiv preprint arXiv:2206.07682}, 2022.

\bibitem[\protect\citeauthoryear{Wu and Raghavendra}{2022}]{WuRaghavendraGupta2022}
Carole~J. Wu and Ramya~others Raghavendra.
\newblock Sustainable ai: Environmental implications, challenges and opportunities.
\newblock In {\em MLSys}, pages 795--813, 2022.

\bibitem[\protect\citeauthoryear{Wu \bgroup \em et al.\egroup }{2022}]{wu2022wav2seq}
Felix Wu, Kwangyoun Kim, et~al.
\newblock Wav2seq: Pre-training speech-to-text encoder-decoder models using pseudo languages.
\newblock {\em arXiv preprint arXiv:2205.01086}, 2022.

\bibitem[\protect\citeauthoryear{Wynsberghe}{2021}]{vanWynsberghe2021}
Almee~V Wynsberghe.
\newblock Sustainable ai: Ai for sustainability and the sustainability of ai.
\newblock {\em AI Ethics}, 1:213--218, 2021.

\bibitem[\protect\citeauthoryear{Xin \bgroup \em et al.\egroup }{2023}]{cheng2023lift}
Cheng Xin, Luo Di, et~al.
\newblock Lift yourself up: Retrieval-augmented text generation with self memory.
\newblock {\em arXiv preprint arXiv:2305.02437}, 2023.

\bibitem[\protect\citeauthoryear{Yang \bgroup \em et al.\egroup }{2023}]{yang2023vid2seq}
Antoine Yang, Arsha Nagrani, et~al.
\newblock Vid2seq: Large-scale pretraining of a visual language model for dense video captioning.
\newblock {\em arXiv preprint arXiv:2302.14115}, 2023.

\bibitem[\protect\citeauthoryear{Yao \bgroup \em et al.\egroup }{2023}]{yao2023editing}
Yunzhi Yao, Peng Wang, et~al.
\newblock Editing large language models: Problems, methods, and opportunities.
\newblock {\em arXiv preprint arXiv:2305.13172}, 2023.

\bibitem[\protect\citeauthoryear{Yosinski \bgroup \em et al.\egroup }{2014}]{yosinski2014transferable}
Jason Yosinski, Jeff Clune, et~al.
\newblock How transferable are features in deep neural networks?
\newblock In {\em NeurIPS}, 2014.

\bibitem[\protect\citeauthoryear{Yu \bgroup \em et al.\egroup }{2023}]{yu2023generate}
Wenhao Yu, Dan Iter, et~al.
\newblock Generate rather than retrieve: Large language models are strong context generators.
\newblock {\em arXiv preprint arXiv:2209.10063}, 2023.

\bibitem[\protect\citeauthoryear{Zaremba \bgroup \em et al.\egroup }{2014}]{zaremba2014recurrent}
Wojciech Zaremba, Ilya Sutskever, et~al.
\newblock Recurrent neural network regularization.
\newblock {\em arXiv preprint arXiv:1409.2329}, 2014.

\bibitem[\protect\citeauthoryear{Zhang \bgroup \em et al.\egroup }{2023a}]{zhang2023texttoimage}
Chenshuang Zhang, Chaoning Zhang, et~al.
\newblock Text-to-image diffusion models in generative ai: A survey.
\newblock {\em arXiv preprint arXiv:2303.07909}, 2023.

\bibitem[\protect\citeauthoryear{Zhang \bgroup \em et al.\egroup }{2023b}]{10337079}
Yi~Zhang, Zhongyang Yu, et~al.
\newblock Long-term memory for large language models through topic-based vector database.
\newblock In {\em IALP}, pages 258--264, 2023.

\end{thebibliography}

\end{document}